# Spin-pump-induced spin transport demonstration in a photoconductive PTCDA molecular thin film with a transparent spin current detector

Yuri Matsukawa, [1] and Eiji Shikoh [1,a)]

AFFILIATIONS

[1]*Graduate School of Engineering, Osaka Metropolitan University, 1-1 Gakuen-cho, Naka-ku, Sakai, 599-8531, Japan*

[a)]Author to whom correspondence should be addressed: shikoh@omu.ac.jp

## ABSTRACT

We demonstrate spin-pump-induced spin transport in a photoconductive PTCDA (3,4,9,10-perylene-teracarboxylic dianhydride) molecular thin film with a transparent ITO ($In_2O_3$ + $SnO_2$) film as a spin current detector. In a tri-layer stacking structure sample composed of ITO/PTCDA/$Ni_{80}Fe_{20}$, pure spin current is generated in the PTCDA layer by using the spin pumping driven by the ferromagnetic resonance of the $Ni_{80}Fe_{20}$ layer. The generated spin current is absorbed into the ITO layer, converted to a charge current due to the inverse spin-Hall effect of the ITO layer, and detected as an electromotive force via the ITO resistance. Also, the light irradiation effect on spin transport in PTCDA films is investigated.

Pure spin current which is a flow of spin angular momenta and a dissipation-less information propagation has attracted much attention since it is expected to be utilized as one of the energy-saving technologies in electronic devices. To control the pure spin current by applying an external field, such as electrical voltage, heat, pressure, light irradiation and so on, is one significant issue for practical use of the pure spin current. Up to now, it is reported that the silicon (Si)-based spin-transistor controlled by applying electrical field has experimentally demonstrated at room temperature (RT).[1] That is, in that demonstration, the spin current property in the Si layer is controlled by changing the carrier density in the Si layer with a so-called gate-voltage application.[1,2] Meanwhile, no successful studies seem to be reported about the spin current control with other external fields.

Organic molecular materials composed of light elements are promising for spin transporting materials because their spin-orbit interaction functioning as spin scattering centers is weak, in general. On spin transport in molecular materials, carrier density, carrier mobility, molecular orientation, hyperfine interaction, and containing elements in the materials are complicatedly affected. However, the relationship between the spin-transport mechanism and those factors in molecular materials has not sufficiently been clarified yet. At present, pure spin current transport in various organic molecular thin films have been experimentally achieved at RT,[3-15] by using a combination method of a dynamical spin injection with the spin pumping driven by the ferromagnetic resonance (FMR),[16,17] and an electrical spin detection with the inverse spin-Hall effect (ISHE),[18,19] and the spin diffusion length ($\lambda$) and/or spin lifetime ($\tau$) in those molecular films have been evaluated.[3-15]

Except for molecular materials with high molecular orientation, in general, organic molecular films tend to have an amorphous state. Two models of spin transport mechanism in such molecular films are suggested:[20] One is due to the hopping transport derived from those low carrier densities (hereafter, called as model (1)), another is due to the exchange coupling between the spins in the case the distance between the nearest neighbor spins is relatively close (hereafter, model (2)). On the model (1), the spin-polarized carriers are diffusively transported in molecular films, and it is known that the $\lambda^2$ is proportional to the $\tau$ even in the disordered molecular films.[3,7,10,13] Meanwhile, the trend between the $\lambda$ and $\tau$ in polymer films with high carrier density prepared with the carrier chemically doping after the film formations [5,8] is different from non-doped polymer films.[3,21] Because the distance between the spins in those carrier-doped polymer films is relatively close due to the increase of the carrier densities, the model (2) would be dominant on the spin transport properties in those carrier-doped films.[5,8] Moreover, it is reported that the spin transport in a nonconjugated radical polymer, where only the situation of the spin transport model (2) exists "in principle," has been achieved. [14] Thus, it is expected that the spin transport properties in the molecular films are operated by controlling the carrier density in molecular films with an external field under an assumption that the carrier density in molecular materials strongly affect spin transport. On the other hand, in small molecular films, some research groups suggest the model (1) is main spin transport mechanism. [6,7,9-13] Meanwhile, it is reported that the model (2) is dominant as the spin transport mechanism in $Alq_3$ molecular films. [4] However, the relationship between the carrier density and the spin transport properties in small molecular films seems not to be investigated, while the relationship between them in polymer films has been investigated. Thus, in this study, the relationship between the carrier density and the spin transport properties in small molecular films is experimentally investigated.

Many kinds of molecular films show photoconductivity, especially with visible light irradiation. That is, those conductive carrier densities in the photoconductive molecular films increased when light having appropriate excitation energy is irradiated. Thinking from this point, if carriers exist in photoconductive molecular films even if the carrier density is very low, to operate pure spin current property in such molecular films with light irradiation might be possible. In the case the model (1) is dominant, the increase of conductive carrier numbers directly affects the spin current density, that is, the spin transport signal in such molecular films will be changed. In the case the model (2) is dominant, the increase of carrier numbers does not directly affect the spin current density. However, the spin transport signal will be changed because the distance between the nearest neighbor spins will be closer in such exciting molecular films and it will cause the strength of the exchange coupling change for spin transport in the films.

In this study, we demonstrate spin-pump-induced spin transport in a photoconductive small molecular thin film with a transparent spin current detector, and the light irradiation effect on the spin transport in the molecular films is investigated, as the first step to achieve the spin transport operation in molecular films with light irradiation. As similar to the previous studies, [3-15] by using a tri-layer structure sample composed of ferromagnetic metal (FM) spin-injector layer/photoconductive molecular layer/non-magnetic spin-detector layer, pure spin current is generated in the photoconductive molecular layer by using the spin pumping driven by the FMR of the FM layer, is absorbed into the non-magnetic layer, converted to a charge current due to the ISHE of the non-magnetic layer, and electrically detected as an EMF via the non-magnetic layer resistance. ITO composed of $In_2O_3$ and $SnO_2$ has been selected as the transparent non-magnetic spin detector. There are two significant reasons why ITO is selected as the spin detector: One is ITO thin film is transparent in visible light region. This means the molecular layer is directly photo-excited via the ITO layer. Another reason is ITO shows the ISHE.[22,23] For a polymer film, there is a spin transport study using the combination method of spin-pumping and ISHE with the ITO spin current detector. [15] However, no light irradiation effect on spin transport in the films has been considered yet. For a small molecular film, there is a spin transport study using the samples composed of FM layer/molecular layer/ITO layer. [24] However, the ISHE of the ITO has not been utilized while the spin-pumping is utilized for the spin injection. In this study, the ISHE of ITO is used as the electrical spin detector for spin transport. Also, (3,4,9,10-perylene-teracarboxylic dianhydride) is selected as a photoconductive molecular material because photoconductivity of PTCDA films [25] and optical property of PTCDA films on ITO films [26] have respectively been clarified.

Figure 1 shows a schematic illustration of our tri-layer stacking structure sample composed of transparent ITO ($In_2O_3$ + $SnO_2$ 10wt%) film/PTCDA film/ferromagnetic $Ni_{80}Fe_{20}$ film, and the evaluation method of the spin transport properties in the PTCDA film. A spin current $\overrightarrow{J_s}$ is generated in the PTCDA layer by using the spin pumping driven by the FMR of the $Ni_{80}Fe_{20}$ layer. The generated $\overrightarrow{J_s}$ is absorbed into the ITO layer, converted to a charge current due to the ISHE of the ITO layer, and detected as an EMF via the ITO resistance.

On a synthesized quartz substrate, ITO film (thickness, 60 nm) was formed by radiofrequency (RF) magnetron sputtering. After pumping the vacuum chamber to < $10^{-4}$ Pa, Ar gas was introduced. During the ITO depositions, the Ar gas pressure was set to 0.80 Pa, and the deposition rate was 0.057 nm/s. Next, a PTCDA film (thickness, $d$ of 30~50 nm) was formed on the ITO film by thermal evaporation via a metal mask. The base pressure was below $10^{-6}$ Pa, and the PTCDA deposition rate was controlled at 0.10 nm/s with monitoring by a quartz resonator. Finally, $Ni_{80}Fe_{20}$ film (thickness, 25 nm) was formed on the PTCDA film by electron beam deposition via another metal mask. The base pressure was below $10^{-6}$ Pa, and the $Ni_{80}Fe_{20}$ deposition rate was controlled at 0.03 nm/s with monitoring by a quartz resonator. During the $Ni_{80}Fe_{20}$ depositions, the sample substrate was cooled under -2 °C to prevent from breaking the molecular films.

Optical transmission and absorbance properties were measured with UV-Vis-NIR spectrophotometer (Shimadzu, UV3600). Luminescence spectrum of our light source (AS ONE,

SLUV-4, or SLUV-8) for light irradiation experiments was measured with a spectrometer (Ocean optics, USB 2000). Electrical properties were evaluated using a two-probe method with a source meter (Keithley Instruments, 2400). To excite FMR in $Ni_{80}Fe_{20}$ films for spin pumping, a coplanar waveguide (CPW) connected to a vector network analyzer (VNA: KEYSIGHT Technology, N5221A) and a couple of electromagnets were used. A nanovoltmeter (Keithley Instruments, 2182A) to detect electromotive forces from samples was used. Leading wires for detecting the output voltage properties were directly attached at both ends of the ITO layer with silver paste. Sample substrates were placed face down on the CPW. In this configuration, light is irradiated to film samples through the quartz substrate on light irradiation experiments. All measurements were performed at RT.

Figure 2(a) shows an optical transmission spectrum of an ITO film with the thickness of 60 nm on a quartz substrate. The transmittance of the ITO film has over 80% in the wavelength range of longer than 360 nm. Figure 2(b) shows an optical absorbance spectrum of a PTCDA film with the *d* of 30 nm on a quartz substrate (red curve) and luminescence spectrum of our light source for light irradiation experiments (black curve). The center wavelength of our light source is 365 nm which almost corresponds to an intrinsic electronic transition (n-$\pi^*$) of PTCDA films [25,26] and the light can be through the ITO film with the transmittance of >80%. That is, in the sample structure shown in Fig. 1, the PTCDA film can be directly excited with our light source irradiation through the quartz substrate and the ITO spin current detector.

Figure 3 shows current voltage properties of a PTCDA thin film with light irradiation (solid curve) and without light irradiation (broken curve). The conductivity of the PTCDA film under the light irradiation is about five times higher than without light irradiation. This is one clear evidence of the photoconductivity of PTCDA films and indicates the conductive carrier density in PTCDA films becomes about five times large with the light irradiation by using our light source under simple Drude model.[27]

Figure 4(a) shows high-frequency electric power transmission properties ($\Delta S_{12}$) corresponding to FMR spectra of our tri-layer samples with the $d$ of 30 nm. $\theta$ is the orientation angle of the static magnetic field ($H$) to the sample film plane. The RF magnetic field frequency ($f$) and power of VNA are set to 5 GHz and 13 dBm (~ 20 mW), respectively. Typical FMR spectra of our tri-layer samples shown in Fig. 1 are observed, and the FMR field ($H_{FMR}$) is 308 Oe. According to the Kittel formula,[28] the FMR condition in an in-plane field is expressed as follows:

$$\frac{\omega}{\gamma} = \sqrt{H_{FMR}(H_{FMR} + 4\pi M_S)}, \qquad (1)$$

where $\omega$, $\gamma$ and $M_S$ are the angular frequency ($2\pi f$), the gyromagnetic ratio for ferromagnetic metal (FM) film, and saturation magnetization of the FM film, respectively. Using eq. (1), the $M_S$ of the $Ni_{80}Fe_{20}$ film in the tri-layer structure sample was estimated to be 709 emu/cc. This $M_S$ value is valid compared to typical $M_S$ of $Ni_{80}Fe_{20}$ films, 700~800 emu/cc.[29]

Fig. 4(b) shows EMF properties of the same sample as used in Fig. 4(a) and simultaneously measured with the data in Fig. 4(a); the circles represent experimental data, and solid lines are fitting curves obtained using the following equation:[18]

$$V(H) = V_{Sym}\frac{\Gamma^2}{(H-H_{FMR})^2+\Gamma^2} + V_{Asym}\frac{-2\Gamma(H-H_{FMR})}{(H-H_{FMR})^2+\Gamma^2}, \qquad (2)$$

where $\Gamma$ denotes damping constant (18 Oe in Fig. 4). The first and second terms in eq. (2) correspond to the symmetry voltage term to $H = H_{FMR}$ due to the ISHE and/or other effects showing the same symmetric voltage behavior, and the asymmetry voltage term to $H = H_{FMR}$ due to the anomalous Hall effect and/or other effects showing the same asymmetric voltage behavior, respectively.[18] $V_{Sym}$ and $V_{Asym}$ correspond to the coefficients of the first and second terms in eq.

(2). In Fig. 4(b), output voltages are observed around $H_{FMR}$ at $\theta$ of 0° and 180°. Notably, the output voltages change their signs between $\theta$ of 0° and 180°. This sign inversion of voltage in ITO associated with the magnetization reversal in $Ni_{80}Fe_{20}$ is one characteristic of ISHE.[18]

Figure 5(a) shows VNA RF power dependence of the EMFs generated in a tri-layer sample with the $d$ of 30 nm under the FMR excitation at the $\theta$ of 0° and the $f$ of 5 GHz. Fig. 5(b) shows RF power dependence of the $V_{sym}$ analyzed from the data in Fig. 5(a) with eq. (2). The $V_{sym}$ is proportional to the applied VNA RF power, which is a characteristic that the $V_{sym}$ originates from the FMR excitation of $Ni_{80}Fe_{20}$ film of our tri-layer structure sample. Those voltage behaviors under the FMR excitation are almost same as the spin transport achievement using Pd as a spin detector.[11] That is, spin transport in a PTCDA thin film was successfully demonstrated even with an ITO spin current detector.

Finally, spin current properties in PTCDA films under light irradiation were investigated. Figure 6 shows experimental examples of light irradiation to a tri-layer sample with the $d$ of 50 nm. The EMFs under the FMR excitation of $Ni_{80}Fe_{20}$ film were measured from (a) to (d) in Fig. 6, in the time order. While (a) and (c) in Fig. 6 are properties without light irradiation, (b) and (d) are properties under light irradiation. For FMR excitation of $Ni_{80}Fe_{20}$ films, the RF field frequency and power of VNA are set to 5 GHz and 13 dBm (~ 20 mW), respectively. $\theta$ is the orientation angle of the $H$ to the sample film plane. The distance from the light source to sample on experiments is 15 cm which is the same as the photoconductivity experiments as shown in Fig. 3. That is, the carrier density in the PTCDA film under the light irradiation might be five times larger than without light irradiation, while an electrical field is applied in Fig. 3. As long as seen in Fig. 6, no clear EMF property changes between with and without light irradiation have been observed yet. As the reasons we haven't observed the light irradiation effects, the following reasons have been considered: Extrinsic reasons are derived from experimental setup. The power of light source is too low and the excitation light wavelength is not optimized, while the photoconductivity in PTCDA films is observed. Those will be improved by using the white light source with higher power, and next, light-wavelength dependent experiments using the white light source equipped with a diffraction grating should be done to optimize. For intrinsic reasons we haven't observed the light irradiation effects, there is possibility that carrier transport and spin transport mechanisms in PTCDA films are completely different. That is, the spin transport model (1) due to carrier hopping transport may be excluded for possible spin transport mechanisms in PTCDA films, although spin transport distance dependence is clearly observed in PTCDA films.[11] However, even if the spin transport model (2) due to exchange coupling is dominant in PTCDA films, electron density dependent spin transport should be observed because electronic states in the films under light irradiation is different from cases without light irradiation because the distance between the nearest neighbor spins in the films under the light irradiation is relatively closer. To verify them, light source improvement for experiments should be done as the first step. Also, other photo-conductive molecular films should be tested near future.

In summary, to aim the clarification of spin transport mechanism in molecular films and the spin current control in the films originating from carrier density control with light irradiation, ITO/PTCDA/$Ni_{80}Fe_{20}$ stacking structure samples were prepared. Spin-pump-induced spin transport in a PTCDA film was successfully demonstrated even with an ITO spin current detector. Also, the light irradiation effect on spin transport in the PTCDA films was investigated.

This research was partly supported by the Grant-in-Aids from the JSPS for Scientific Research (No. 23K04569, 26K08180) and by the Cooperative Research Program of "Network Joint Research Center for Materials and Devices".

## AUTHOR DECLARATIONS

### Conflict of Interest

The authors declare no competing financial interests.

### Author Contributions

Yuri Matsukawa: Conceptualization (equal); All experiments including setups; Data analysis (equal); Discussion (equal); Writing – original draft (equal).

Eiji Shikoh: Conceptualization (equal); Funding acquisition; Project administration; Supervision; Experimental system setup (partly support); Data analysis (equal); Discussion (equal); Writing – original draft (equal); Editing.

## DATA AVAILABILITY

The data that supports the findings of this study are available from the corresponding author upon reasonable request.

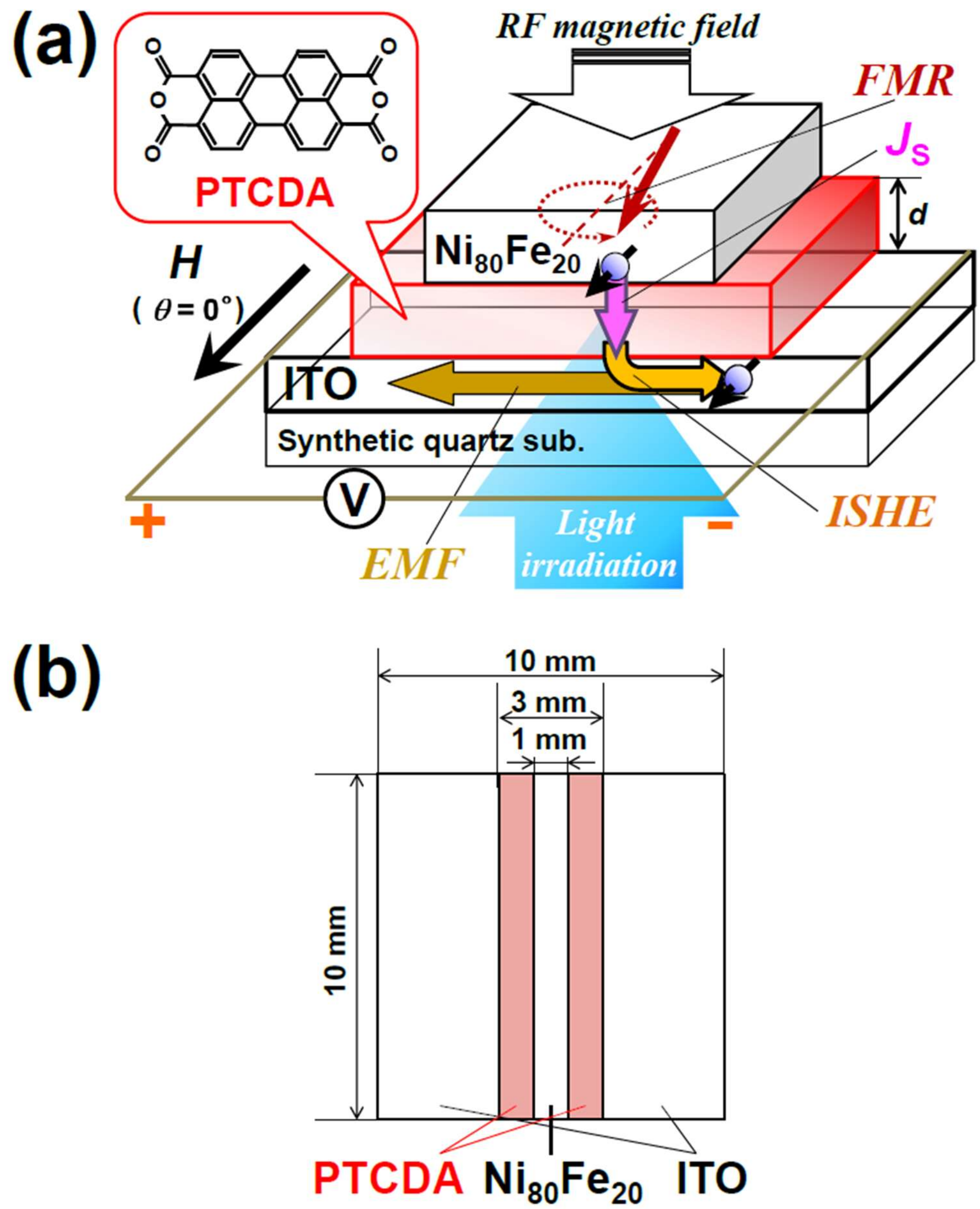

FIG. 1. (a) Bird's-eye-view and (b) top-view illustrations of our sample and experimental setup. $H$, $J_S$, and *EMF* correspond to applied static magnetic field, spin current generated in a PTCDA film by spin-pumping, and electromotive force due to the ISHE in the ITO film, respectively. $\theta$ is the orientation angle of the $H$ to the sample film plane.

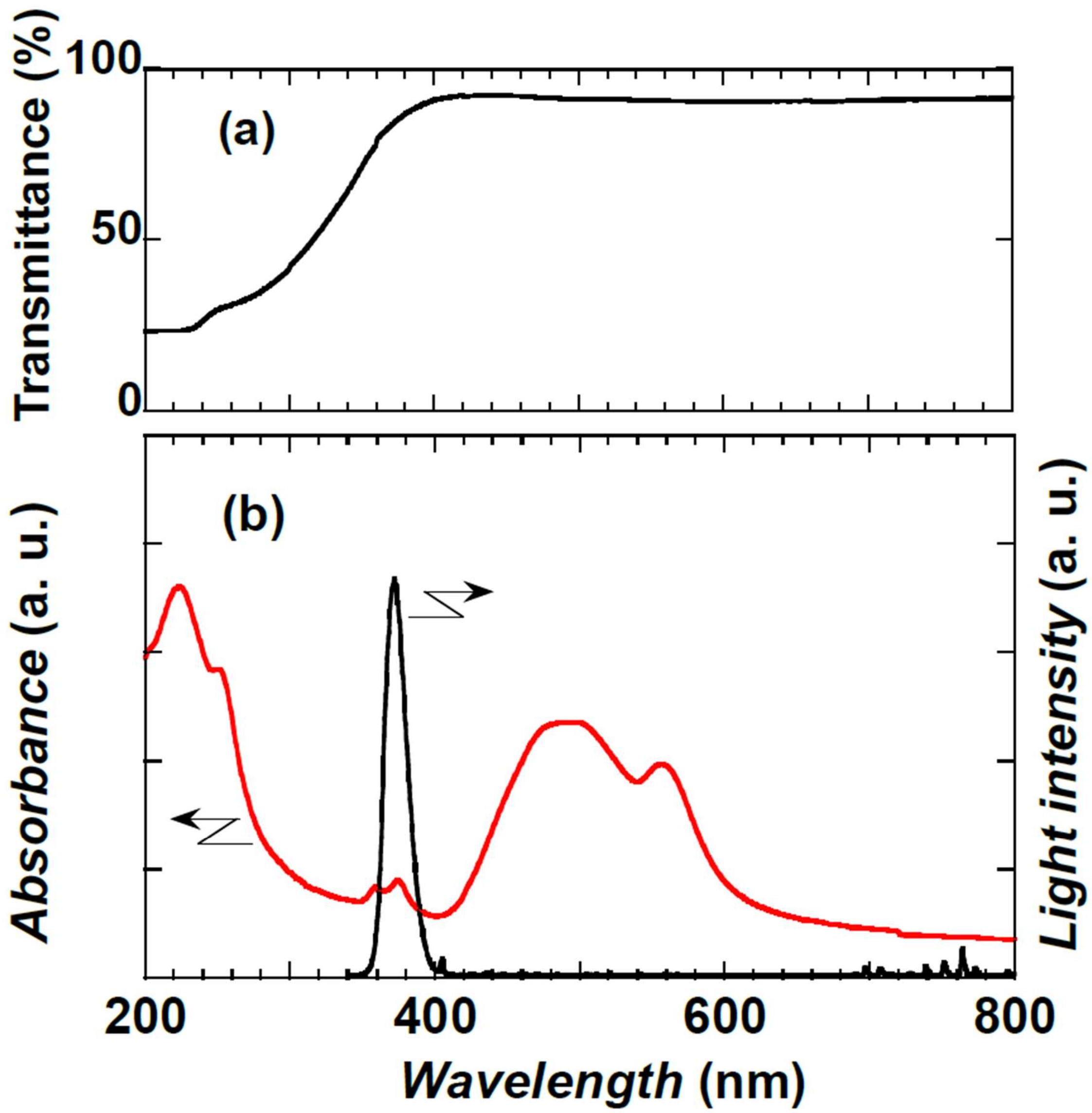

FIG. 2. (a) Optical transmission spectrum of an ITO film with the thickness of 60 nm on a quartz substrate. (b) Optical absorbance spectrum of a PTCDA film with the thickness of 30 nm on a quartz substrate (red curve) and luminescence spectrum of our light source for light irradiation experiments (black curve).

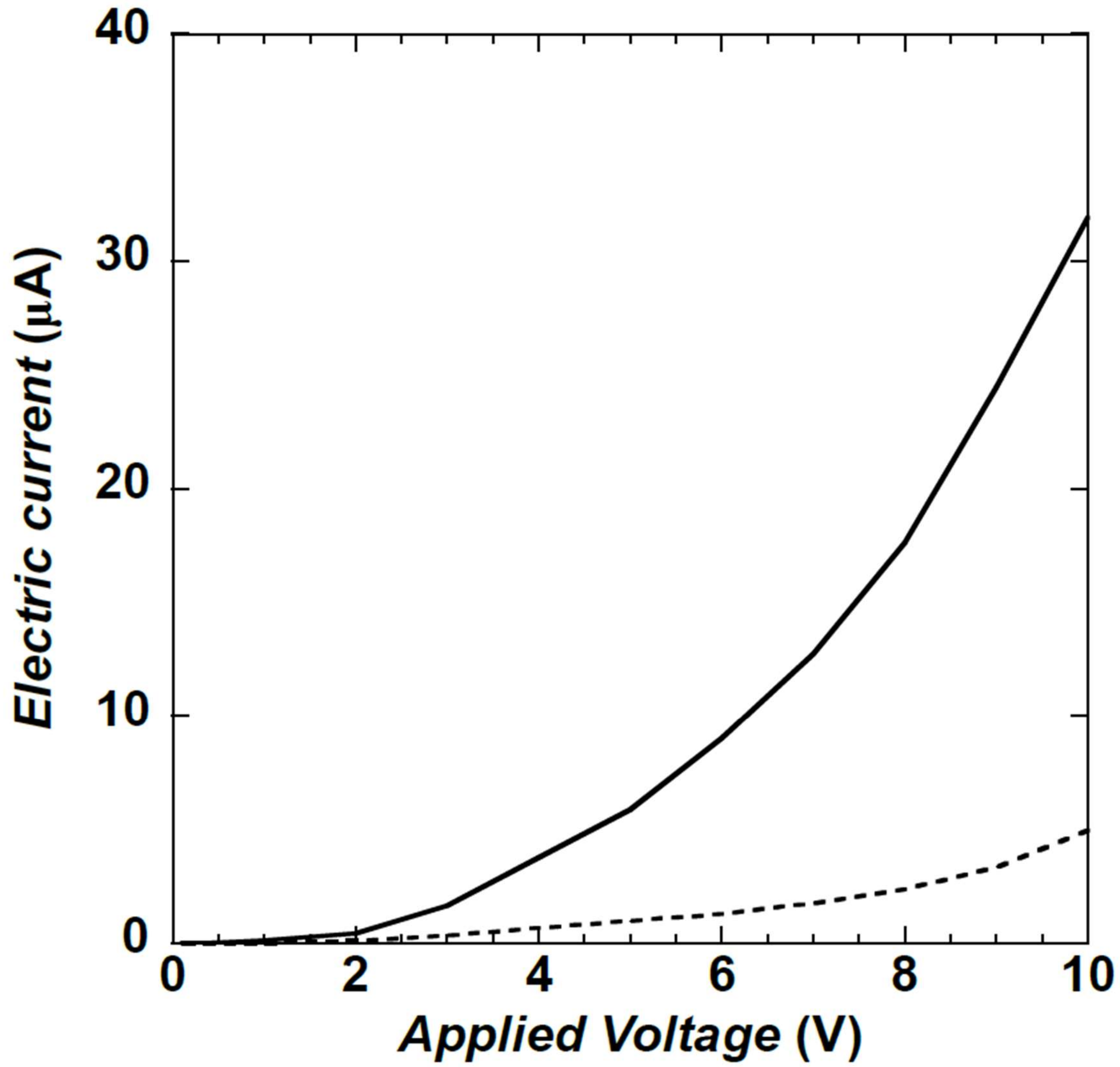

FIG. 3. Current voltage properties of a PTCDA thin film with light irradiation (solid curve) and without light irradiation (broken curve).

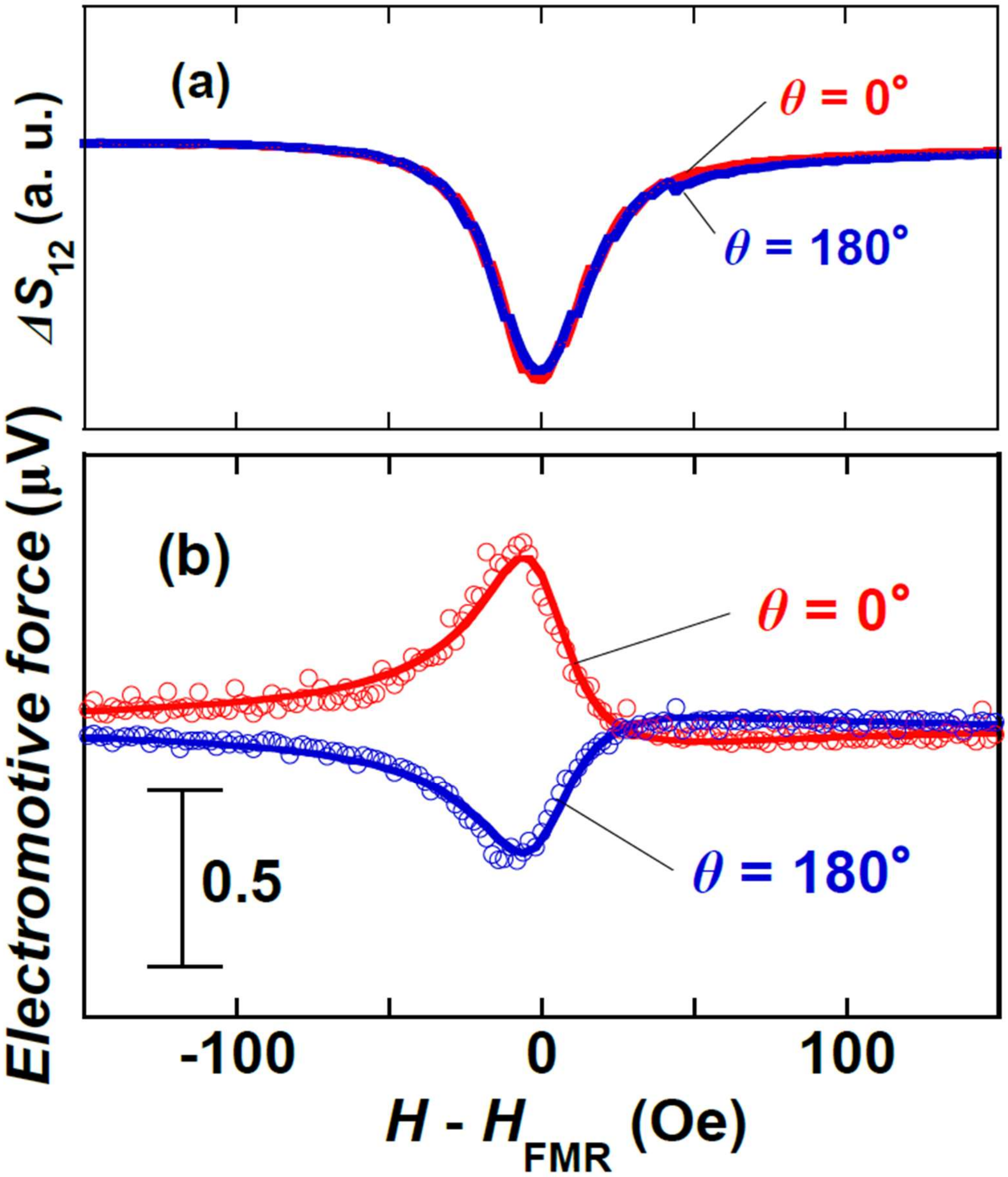

FIG. 4. (a) Typical FMR spectra for our tri-layer sample, and (b) electromotive force properties generated in the same tri-layer sample under the FMR excitations. The RF field frequency and power are set to 5 GHz and 13 dBm (~ 20 mW), respectively. $\theta$ is the orientation angle of the $H$ to the sample film plane. Solid curves in (b) are fittings with eq. (2).

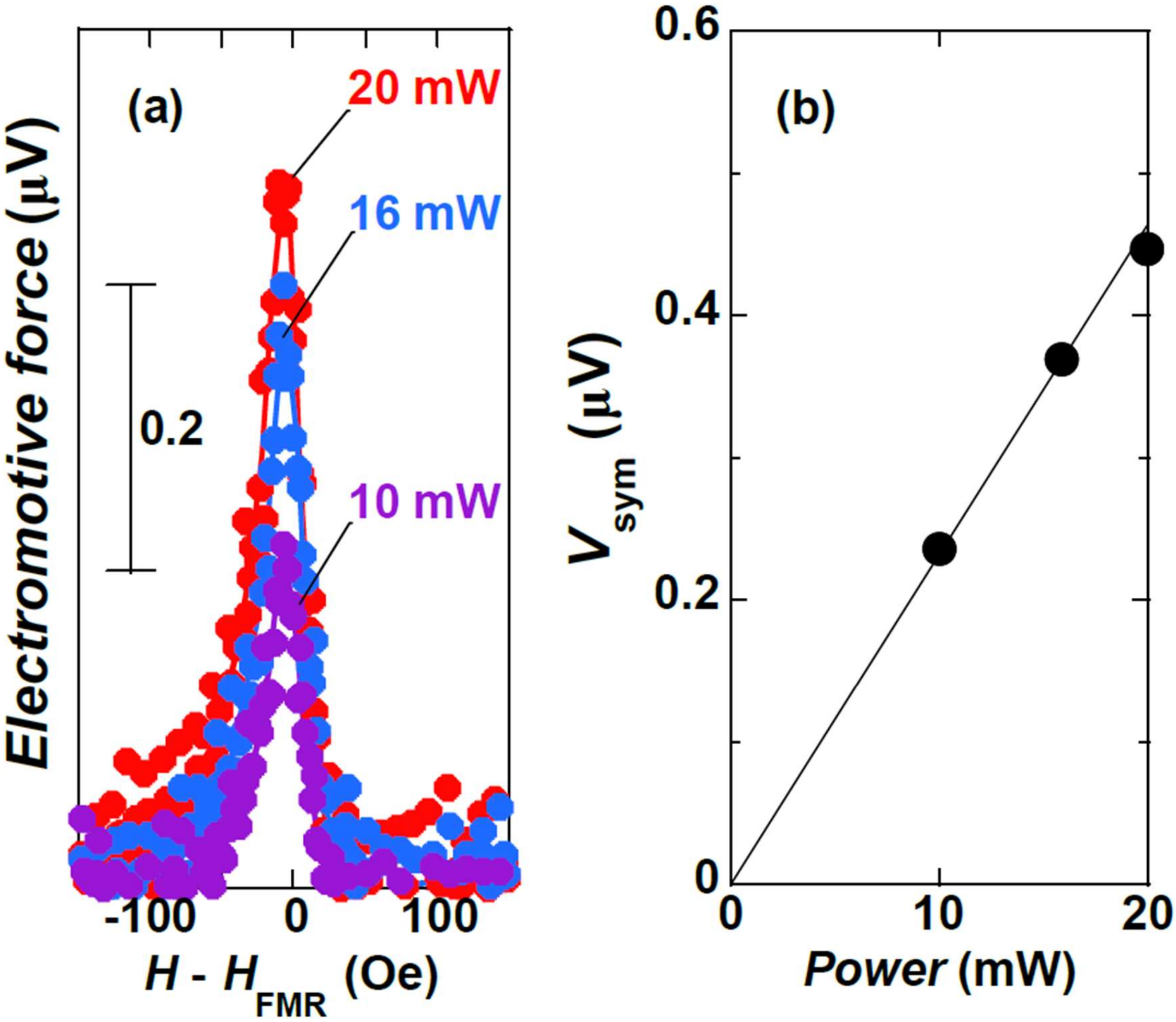

FIG. 5. (a) VNA RF power dependence of the electromotive forces generated in a tri-layer sample under the FMR excitation with the orientation angle of the $H$ to the sample film plane, $\theta$ is 0. The RF field frequency is set to 5 GHz. (b) VNA RF power dependence of the $V_{sym}$ analyzed from the data in (a) with eq. (2).

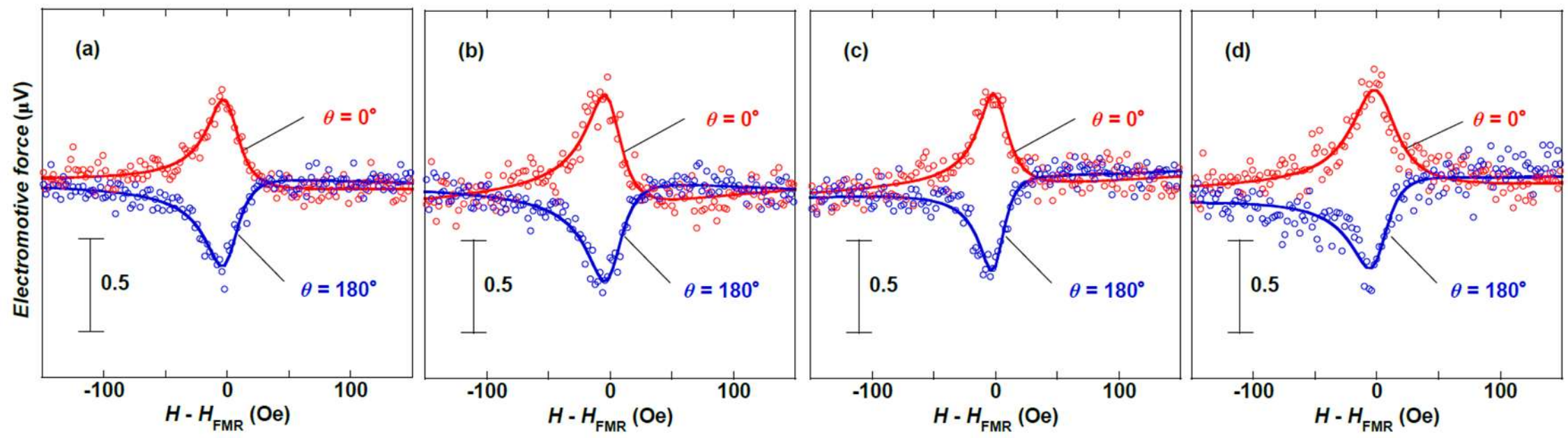

FIG. 6. Light irradiation effects to a tri-layer sample. The electromotive force properties under the FMR of the $Ni_{80}Fe_{20}$ layer were measured from (a) to (d) in the time order. While (a) and (c) are properties without light irradiation, (b) and (d) are properties under light irradiation. The RF field frequency and power are set to 5 GHz and 13 dBm (~ 20 mW), respectively. $\theta$ is the orientation angle of the $H$ to the sample film plane. Solid curves are fittings with eq. (2).